\documentstyle[twoside,fleqn,espcrc2,epsfig,amsmath,amssymb,feynarts]{article}

\makeatletter
\def\citer{\@ifnextchar [{\@tempswatrue\@citexr}{\@tempswafalse\@citexr[]}}
 
\def\@citexr[#1]#2{\if@filesw\immediate\write\@auxout{\string\citation{#2}}\fi
  \def\@citea{}\@cite{\@for\@citeb:=#2\do
    {\@citea\def\@citea{--\penalty\@m}\@ifundefined
       {b@\@citeb}{{\bf ?}\@warning
       {Citation `\@citeb' on page \thepage \space undefined}}%
\hbox{\csname b@\@citeb\endcsname}}}{#1}}
\makeatother

\def\refeq#1{\mbox{Eq.~(\ref{#1})}}

\def\reffi#1{\mbox{Fig.~\ref{#1}}}
\def\reffis#1{\mbox{Figs.~\ref{#1}}}

\def\citere#1{\mbox{Ref.~\cite{#1}}}

\newcommand{\ie}{i.e.\ }

\def\xtilde#1{%
  \setbox0\hbox{$\tilde#1$}%
  \rlap{\raise\ht0\hbox{\tiny$_{\,(\;\,)}$}}%
  \tilde#1%
}

\newcommand{\At}{A_t}
\newcommand{\Ab}{A_b}

\newcommand{\Xt}{X_t}

\newcommand{\msusy}{M_{\text{SUSY}}}

\newcommand{\msbarm}{\overline{\rm{MS}}}

\def\order#1{${\cal O}(#1)$}
\newcommand{\cp}{{\cal CP}}

\newcommand{\edz}{\frac{1}{2}}
\def\ed#1{\frac{1}{#1}}

\newcommand{\onel}{one-loop}

\newcommand{\fh}{{\em FeynHiggs}}

\newcommand{\MW}{M_W}
\newcommand{\MZ}{M_Z}
\newcommand{\MA}{M_A}
\newcommand{\mh}{m_h}
\newcommand{\mH}{m_H}

\newcommand{\MH}{M_H}

\newcommand{\MHp}{M_{H^\pm}}
\newcommand{\mhmax}{m_h^{\rm max}}
\newcommand{\siHenh}{\si_H^{\text{enh}}}

\newcommand{\mt}{m_{t}}

\newcommand{\mgl}{m_{\tilde{g}}}

\newcommand{\tsf}{\theta\kern-.20em_{\tilde{f}}}
\newcommand{\tsfp}{\theta\kern-.20em_{\tilde{f}\prime}}
\newcommand{\tsq}{\theta\kern-.15em_{\tilde{q}}}
\newcommand{\sw}{s_W}
\newcommand{\cw}{c_W}

\newcommand{\KL}{\left(}
\newcommand{\KR}{\right)}
\newcommand{\KKL}{\left[}
\newcommand{\KKR}{\right]}

\newcommand{\VL}{\left( \begin{array}{c}}
\newcommand{\VR}{\end{array} \right)}
\newcommand{\ML}{\left( \begin{array}{cc}}
\newcommand{\MLd}{\left( \begin{array}{ccc}}
\newcommand{\MLv}{\left( \begin{array}{cccc}}
\newcommand{\MR}{\end{array} \right)}

\newcommand{\re}{\mathop{\rm Re}}

\newcommand{\tb}{\tan \beta}

\newcommand{\sinb}{\sin \beta\hspace{1mm}}

\newcommand{\Cb}{\cos \beta\hspace{1mm}}

\newcommand{\Sa}{\sin \alpha\hspace{1mm}}

\newcommand{\Ca}{\cos \alpha\hspace{1mm}}

\newcommand{\Sba}{\sin (\beta - \alpha)}
\newcommand{\Cba}{\cos (\beta - \alpha)}

\newcommand{\CZa}{\cos 2\alpha\hspace{1mm}}

\newcommand{\tev}{\,\, {\rm TeV}}
\newcommand{\gev}{\,\, {\rm GeV}}

\newcommand{\BC}{\begin{center}}
\newcommand{\EC}{\end{center}}
\newcommand{\BE}{\begin{equation}}
\newcommand{\EE}{\end{equation}}
\newcommand{\BEA}{\begin{eqnarray}}
\newcommand{\BEAnn}{\begin{eqnarray*}}
\newcommand{\EEA}{\end{eqnarray}}
\newcommand{\EEAnn}{\end{eqnarray*}}
\newcommand{\non}{\nonumber}
\newcommand{\id}{{\rm 1\kern-.12em
\rule{0.3pt}{1.5ex}\raisebox{0.0ex}{\rule{0.1em}{0.3pt}}}}
\newcommand{\lsim}
{\;\raisebox{-.3em}{$\stackrel{\displaystyle <}{\sim}$}\;}
\newcommand{\gsim}
{\;\raisebox{-.3em}{$\stackrel{\displaystyle >}{\sim}$}\;}

\def\al{\alpha}
\def\aeff{\al_{\rm eff}}

\def\be{\beta}

\def\ga{\gamma}

\def\de{\delta}

\def\si{\sigma}

\def\Ga{\Gamma}

\def\Ghn{\Ga_h^{(0)}}
\def\GHn{\Ga_H^{(0)}}

\def\De{\Delta}

\def\Si{\Sigma}

\newcommand{\eennH}{$e^+e^- \to \bar\nu \nu \, H$}

\newcommand{\fb}{\mbox{~fb}}

\newcommand{\iabm}{\mbox{ab}^{-1}}

\newcommand{\WWH}{WW \to H}

\newcommand{\rZH}{\hat Z_H}

\newcommand{\rZhH}{\hat Z_{hH}}

\allowdisplaybreaks

\begin{document}

\thispagestyle{empty}
\setcounter{page}{0}
\def\thefootnote{\fnsymbol{footnote}}

\begin{flushright}
DCPT/02/124\\ 
IPPP/02/62\\
LMU 13/02\\
MPI-PhT/2002-72\\
hep-ph/0211384
\end{flushright}

\vspace{1cm}

\begin{center}

{\large\sc {\bf Very Heavy MSSM Higgs-Boson Production at the Linear Collider}}
\footnote{talk given by S.~Heinemeyer at the ``RADCOR 2002 -- Loops \&
Legs 2002'', Kloster Banz, Germany, September 2002}

\vspace{1cm}

{\sc 
T.~Hahn$^{1}$%
\footnote{email: hahn@feynarts.de}%
, S.~Heinemeyer$^{2}$%
\footnote{email: Sven.Heinemeyer@physik.uni-muenchen.de}%
, and G.~Weiglein$^{3}$%
\footnote{email: Georg.Weiglein@durham.ac.uk}
}
 
\vspace{1cm}

{\sl
$^1$Max-Planck-Institut f\"ur Physik (Werner-Heisenberg-Institut),
F\"ohringer Ring 6, \\
D--80805 Munich, Germany

\vspace*{0.4cm}

$^2$Institut f\"ur Theoretische Elementarteilchenphysik,
LMU M\"unchen, Theresienstr.\ 37, D--80333 Munich, Germany

\vspace*{0.4cm}

$^3$Institute for Particle Physics Phenomenology, University of Durham,\\
Durham DH1~3LE, UK

}

\end{center}

\BC
{\bf Abstract}
\EC
In the Minimal Supersymmetric Standard Model (MSSM) we present the
corrections to the heavy neutral $\cp$-even Higgs-boson production in the
$WW$-fusion and Higgs-strahlung channel, \eennH, taking into account all
\order{\al} corrections arising from loops of fermions and sfermions.  
While the $H$~boson
shows decoupling behavior at the tree-level, we find non-negligible
loop corrections that can enhance the cross section considerably.
At a center-of-mass energy of $\sqrt{s} = 1000 \gev$, masses of up to
$\MH \lsim 750 \gev$ are accessible at the LC 
in favorable regions of the MSSM parameter space.

\def\thefootnote{\arabic{footnote}}
\setcounter{footnote}{0}

\newpage


\title{Very Heavy MSSM Higgs-Boson Production at the Linear Collider}

\author{T. Hahn\address{Max-Planck-Institut f\"ur Physik 
(Werner-Heisenberg-Institut), F\"ohringer Ring 6, \\
D--80805 Munich, Germany}%
, S. Heinemeyer\address{Institut f\"ur Theoretische Elementarteilchenphysik,
LMU M\"unchen, Theresienstr.\ 37,\\ D--80333 Munich, Germany}
and
G. Weiglein\address{Institute for Particle Physics Phenomenology, 
University of Durham, Durham DH1~3LE, UK}}

\begin{abstract}
In the Minimal Supersymmetric Standard Model (MSSM) we present the
corrections to the heavy neutral $\cp$-even Higgs-boson production in the
$WW$-fusion and Higgs-strahlung channel, \eennH, taking into account all
\order{\al} corrections arising from loops of fermions and sfermions.  
While the $H$~boson
shows decoupling behavior at the tree-level, we find non-negligible
loop corrections that can enhance the cross section considerably.
At a center-of-mass energy of $\sqrt{s} = 1000 \gev$, masses of up to
$\MH \lsim 750 \gev$ are accessible at the LC 
in favorable regions of the MSSM parameter space.
\end{abstract}

\maketitle


\section{INTRODUCTION}

Finding the mechanism that controls
 electroweak symmetry breaking is one of
the main tasks of the current and next generation of colliders. 
The solution may be the Higgs mechanism within the Standard Model
 (SM), or within its most appealing extension, the Minimal
 Supersymmetric Standard Model (MSSM)~\cite{susy}.
Contrary to the SM, two Higgs doublets are
required in the MSSM, 
resulting in five physical Higgs bosons~\cite{hhg}.
While the discovery of one light
Higgs boson might well be compatible with the predictions both of the SM
and the MSSM, the discovery of an additional heavy Higgs boson would 
be a clear signal for physics beyond the SM. 

The Higgs sector of the MSSM can be expressed at lowest order in terms
 of $\MZ$, $\MA$  
(the mass of the $\cp$-odd Higgs boson), and $\tb = v_2/v_1$, the ratio
of the two vacuum expectation values. 
In the decoupling limit, \ie for $\MA\gsim 200 \gev$, the heavy MSSM
Higgs bosons are nearly degenerate in mass, $\MA \approx \MH \approx \MHp$.
The couplings of the heavy, neutral Higgs bosons to SM gauge bosons
are proportional to $(V = Z, W^\pm)$
\begin{equation}
VV\{h, H\} \sim VA\{H,h\} \sim \{\sin,\cos\}(\be-\al) ,
\end{equation}
where $\al$ is the angle that 
diagonalizes the $\cp$-even Higgs sector.
In the decoupling limit one finds $\be - \al \to
\pi/2$, \ie 
$\{\sin,\cos\}(\be-\al) \to \{1,0\}$.

At the LC, the possible channels for heavy, neutral Higgs-boson
production are the production via $Z$-boson exchange,
\begin{equation}
\begin{aligned}
e^+e^- &\to Z^* \to \{Z,A\}H 
\end{aligned}
\end{equation}
and the $WW$-fusion channel,
\begin{equation}
e^+e^- \to \bar\nu_e W^+ \; \nu_e W^- \to \bar\nu_e \nu_e H \,.
\label{eq:eennhH}
\end{equation}
As a consequence of the coupling structure, in the decoupling limit
the heavy Higgs boson can only be produced in $(H,A)$~pairs.
This limits the LC reach to $\MH\lsim\sqrt s/2$.
Higher-order corrections to the $\WWH$ channel from loops of fermions
and sfermions, however, involve potentially large contributions from the
top and bottom Yukawa couplings and can thus significantly affect the
decoupling behavior. The same corrections may also contribute to the
Higgs-strahlung channel, $e^+e^- \to Z^* \to ZH$. However, while this
channel is suppressed with $1/s$, the $WW$-fusion channels rises with
$\log s$.

\begin{figure}[ht!]
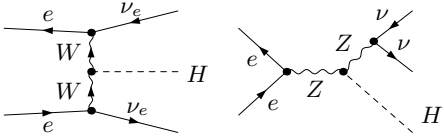

\vspace{-4em}
\begin{center}
\begin{small}
\unitlength=1.0bp%
\begin{feynartspicture}(216,72)(2,1)
\FADiagram{}
\FAProp(0.,15.)(10.,14.5)(0.,){/Straight}{-1}
\FALabel(5.0774,15.8181)[b]{$e$}
\FAProp(0.,5.)(10.,5.5)(0.,){/Straight}{1}
\FALabel(5.0774,4.18193)[t]{$e$}
\FAProp(20.,17.)(10.,14.5)(0.,){/Straight}{1}
\FALabel(14.6241,16.7737)[b]{$\nu_e$}
\FAProp(20.,10.)(10.,10.)(0.,){/ScalarDash}{0}
\FALabel(21.,8.82)[lb]{$H$}
\FAProp(20.,3.)(10.,5.5)(0.,){/Straight}{-1}
\FALabel(15.3759,5.27372)[b]{$\nu_e$}
\FAProp(10.,14.5)(10.,10.)(0.,){/Sine}{-1}
\FALabel(8.93,12.25)[r]{$W$}
\FAProp(10.,5.5)(10.,10.)(0.,){/Sine}{1}
\FALabel(8.93,7.75)[r]{$W$}
\FAVert(10.,14.5){0}
\FAVert(10.,5.5){0}
\FAVert(10.,10.){0}

\FADiagram{}
\FAProp(5.,15.)(10.5,10.)(0.,){/Straight}{-1}
\FALabel(7.18736,11.8331)[tr]{$e$}
\FAProp(5.,5.)(10.5,10.)(0.,){/Straight}{1}
\FALabel(8.31264,6.83309)[tl]{$e$}
\FAProp(25.,17.)(20.5,13.5)(0.,){/Straight}{1}
\FALabel(22.2784,15.9935)[br]{$\nu$}
\FAProp(25.,10.)(20.5,13.5)(0.,){/Straight}{-1}
\FALabel(23.2216,12.4935)[bl]{$\nu$}
\FAProp(25.,3.)(17.,10.)(0.,){/ScalarDash}{0}
\FALabel(28.5,6.00165)[tr]{$H$}
\FAProp(10.5,10.)(17.,10.)(0.,){/Sine}{0}
\FALabel(13.75,8.93)[t]{$Z$}
\FAProp(20.5,13.5)(17.,10.)(0.,){/Sine}{0}
\FALabel(18.134,12.366)[br]{$Z$}
\FAVert(10.5,10.){0}
\FAVert(20.5,13.5){0}
\FAVert(17.,10.){0}
\end{feynartspicture}
\end{small}
\vspace{-5em}
\caption{%
The tree-level diagrams for the process \eennH, consisting of the
$WW$-fusion contribution (left) and the Higgs-strahlung contribution
(right). 
}
\label{fig:tree}
\end{center}
\vspace{-3em}
\end{figure}

Electroweak loop effects on processes within the MSSM where a single Higgs 
boson is produced have recently drawn considerable interest in the
literature~\cite{eennH,Wiener,eennA,eehZhA,logan&su,WH_obr}, 
see \citere{eennH} for a detailed overview, also including
the SM case.
It has been found that the processes
$e^+e^- \to \nu_e \bar \nu_e A$~\cite{eennA}, 
$e^+e^- \to Z^* \to H\{Z,A\}$~\cite{eehZhA}, 
and $e^+e^- \to W^+H^-$~\cite{logan&su,WH_obr} 
only possess a small potential to produce the heavy Higgs bosons with 
$\MH \approx \MA \approx \MHp > \sqrt{s}/2$.
Recently, the results for the one-loop corrections of 
fermions and sfermions to the process \eennH\ have been
presented~\cite{eennH}. 
This process can be mediated via the $WW$-fusion and the 
Higgs-strahlung mechanism, see \reffi{fig:tree}. In the latter case
the $Z$~boson 
is connected to a neutrino pair, $e^+e^- \to ZH \to \bar\nu_l\nu_l H$, 
with $l = e, \mu, \tau$ (where the two latter neutrinos
result in an indistinguishable final state in the detector). 
While the well-known universal Higgs-boson propagator corrections turned
out not to significantly modify the decoupling behavior of the heavy 
$\cp$-even Higgs boson, an analysis of the process-specific
contributions to the $WWH$ vertex has been missing so far. 
We summarize in this paper the LC reach for the heavy $\cp$-even
Higgs boson, including also the effects of beam polarization in our
analysis.

\section{THE CALCULATION}

\begin{figure}[ht!]
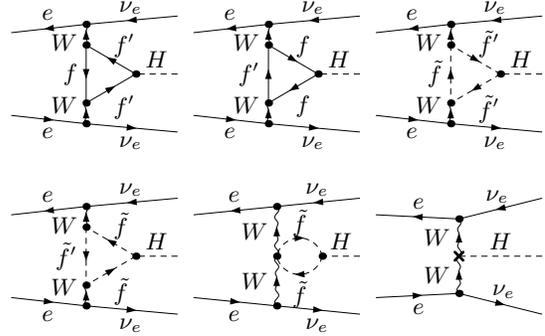

\vspace{-3em}
\begin{center}
\begin{small}
\unitlength=.68bp%
\begin{feynartspicture}(324,202)(3,2)
\FADiagram{}
\FAProp(0.,15.)(9.,16.)(0.,){/Straight}{-1}
\FALabel(4.32883,16.5605)[b]{$e$}
\FAProp(0.,5.)(9.,4.)(0.,){/Straight}{1}
\FALabel(4.32883,3.43948)[t]{$e$}
\FAProp(20.,17.)(9.,16.)(0.,){/Straight}{1}
\FALabel(14.3597,17.5636)[b]{$\nu_e$}
\FAProp(20.,10.)(15.,10.)(0.,){/ScalarDash}{0}
\FALabel(17.5,10.82)[b]{$H$}
\FAProp(20.,3.)(9.,4.)(0.,){/Straight}{-1}
\FALabel(14.3597,2.43637)[t]{$\nu_e$}
\FAProp(9.,16.)(9.,13.5)(0.,){/Sine}{-1}
\FALabel(7.93,14.75)[tr]{$W$}
\FAProp(9.,4.)(9.,6.5)(0.,){/Sine}{1}
\FALabel(7.93,5.25)[br]{$W$}
\FAProp(15.,10.)(9.,13.5)(0.,){/Straight}{1}
\FALabel(12.301,12.6089)[bl]{$f'$}
\FAProp(15.,10.)(9.,6.5)(0.,){/Straight}{-1}
\FALabel(12.301,7.39114)[tl]{$f'$}
\FAProp(9.,13.5)(9.,6.5)(0.,){/Straight}{1}
\FALabel(7.93,10.)[r]{$f$}
\FAVert(9.,16.){0}
\FAVert(9.,4.){0}
\FAVert(15.,10.){0}
\FAVert(9.,13.5){0}
\FAVert(9.,6.5){0}

\FADiagram{}
\FAProp(0.,15.)(9.,16.)(0.,){/Straight}{-1}
\FALabel(4.32883,16.5605)[b]{$e$}
\FAProp(0.,5.)(9.,4.)(0.,){/Straight}{1}
\FALabel(4.32883,3.43948)[t]{$e$}
\FAProp(20.,17.)(9.,16.)(0.,){/Straight}{1}
\FALabel(14.3597,17.5636)[b]{$\nu_e$}
\FAProp(20.,10.)(15.,10.)(0.,){/ScalarDash}{0}
\FALabel(17.5,10.82)[b]{$H$}
\FAProp(20.,3.)(9.,4.)(0.,){/Straight}{-1}
\FALabel(14.3597,2.43637)[t]{$\nu_e$}
\FAProp(9.,16.)(9.,13.5)(0.,){/Sine}{-1}
\FALabel(7.93,14.75)[tr]{$W$}
\FAProp(9.,4.)(9.,6.5)(0.,){/Sine}{1}
\FALabel(7.93,5.25)[br]{$W$}
\FAProp(15.,10.)(9.,13.5)(0.,){/Straight}{-1}
\FALabel(12.301,12.6089)[bl]{$f$}
\FAProp(15.,10.)(9.,6.5)(0.,){/Straight}{1}
\FALabel(12.301,7.39114)[tl]{$f$}
\FAProp(9.,13.5)(9.,6.5)(0.,){/Straight}{-1}
\FALabel(7.93,10.)[r]{$f'$}
\FAVert(9.,16.){0}
\FAVert(9.,4.){0}
\FAVert(15.,10.){0}
\FAVert(9.,13.5){0}
\FAVert(9.,6.5){0}

\FADiagram{}
\FAProp(0.,15.)(9.,16.)(0.,){/Straight}{-1}
\FALabel(4.32883,16.5605)[b]{$e$}
\FAProp(0.,5.)(9.,4.)(0.,){/Straight}{1}
\FALabel(4.32883,3.43948)[t]{$e$}
\FAProp(20.,17.)(9.,16.)(0.,){/Straight}{1}
\FALabel(14.3597,17.5636)[b]{$\nu_e$}
\FAProp(20.,10.)(15.,10.)(0.,){/ScalarDash}{0}
\FALabel(17.5,10.82)[b]{$H$}
\FAProp(20.,3.)(9.,4.)(0.,){/Straight}{-1}
\FALabel(14.3597,2.43637)[t]{$\nu_e$}
\FAProp(9.,16.)(9.,13.5)(0.,){/Sine}{-1}
\FALabel(7.93,14.75)[tr]{$W$}
\FAProp(9.,4.)(9.,6.5)(0.,){/Sine}{1}
\FALabel(7.93,5.25)[br]{$W$}
\FAProp(15.,10.)(9.,13.5)(0.,){/ScalarDash}{-1}
\FALabel(12.301,12.6089)[bl]{$\tilde f'$}
\FAProp(15.,10.)(9.,6.5)(0.,){/ScalarDash}{1}
\FALabel(12.301,7.39114)[tl]{$\tilde f'$}
\FAProp(9.,13.5)(9.,6.5)(0.,){/ScalarDash}{-1}
\FALabel(7.93,10.)[r]{$\tilde f$}
\FAVert(9.,16.){0}
\FAVert(9.,4.){0}
\FAVert(15.,10.){0}
\FAVert(9.,13.5){0}
\FAVert(9.,6.5){0}

\FADiagram{}
\FAProp(0.,15.)(9.,16.)(0.,){/Straight}{-1}
\FALabel(4.32883,16.5605)[b]{$e$}
\FAProp(0.,5.)(9.,4.)(0.,){/Straight}{1}
\FALabel(4.32883,3.43948)[t]{$e$}
\FAProp(20.,17.)(9.,16.)(0.,){/Straight}{1}
\FALabel(14.3597,17.5636)[b]{$\nu_e$}
\FAProp(20.,10.)(15.,10.)(0.,){/ScalarDash}{0}
\FALabel(17.5,10.82)[b]{$H$}
\FAProp(20.,3.)(9.,4.)(0.,){/Straight}{-1}
\FALabel(14.3597,2.43637)[t]{$\nu_e$}
\FAProp(9.,16.)(9.,13.5)(0.,){/Sine}{-1}
\FALabel(7.93,14.75)[tr]{$W$}
\FAProp(9.,4.)(9.,6.5)(0.,){/Sine}{1}
\FALabel(7.93,5.25)[br]{$W$}
\FAProp(15.,10.)(9.,13.5)(0.,){/ScalarDash}{1}
\FALabel(12.301,12.6089)[bl]{$\tilde f$}
\FAProp(15.,10.)(9.,6.5)(0.,){/ScalarDash}{-1}
\FALabel(12.301,7.39114)[tl]{$\tilde f$}
\FAProp(9.,13.5)(9.,6.5)(0.,){/ScalarDash}{1}
\FALabel(7.93,10.)[r]{$\tilde f'$}
\FAVert(9.,16.){0}
\FAVert(9.,4.){0}
\FAVert(15.,10.){0}
\FAVert(9.,13.5){0}
\FAVert(9.,6.5){0}

\FADiagram{}
\FAProp(0.,15.)(10.,16.)(0.,){/Straight}{-1}
\FALabel(4.84577,16.5623)[b]{$e$}
\FAProp(0.,5.)(10.,4.)(0.,){/Straight}{1}
\FALabel(4.84577,3.43769)[t]{$e$}
\FAProp(20.,17.)(10.,16.)(0.,){/Straight}{1}
\FALabel(14.8458,17.5623)[b]{$\nu_e$}
\FAProp(20.,10.)(15.5,10.)(0.,){/ScalarDash}{0}
\FALabel(17.75,10.82)[b]{$H$}
\FAProp(20.,3.)(10.,4.)(0.,){/Straight}{-1}
\FALabel(14.8458,2.43769)[t]{$\nu_e$}
\FAProp(10.,16.)(10.,10.)(0.,){/Sine}{-1}
\FALabel(8.93,13.)[r]{$W$}
\FAProp(10.,4.)(10.,10.)(0.,){/Sine}{1}
\FALabel(8.93,7.)[r]{$W$}
\FAProp(15.5,10.)(10.,10.)(0.8,){/ScalarDash}{-1}
\FALabel(12.75,13)[b]{$\tilde f$}
\FAProp(15.5,10.)(10.,10.)(-0.8,){/ScalarDash}{1}
\FALabel(12.75,7)[t]{$\tilde f$}
\FAVert(10.,16.){0}
\FAVert(10.,4.){0}
\FAVert(15.5,10.){0}
\FAVert(10.,10.){0}

\FADiagram{}
\FAProp(0.,15.)(10.,14.5)(0.,){/Straight}{-1}
\FALabel(5.0774,15.8181)[b]{$e$}
\FAProp(0.,5.)(10.,5.5)(0.,){/Straight}{1}
\FALabel(5.0774,4.18193)[t]{$e$}
\FAProp(20.,17.)(10.,14.5)(0.,){/Straight}{1}
\FALabel(14.6241,16.7737)[b]{$\nu_e$}
\FAProp(20.,10.)(10.,10.)(0.,){/ScalarDash}{0}
\FALabel(15.,10.82)[b]{$H$}
\FAProp(20.,3.)(10.,5.5)(0.,){/Straight}{-1}
\FALabel(15.3759,5.27372)[b]{$\nu_e$}
\FAProp(10.,14.5)(10.,10.)(0.,){/Sine}{-1}
\FALabel(8.93,12.25)[r]{$W$}
\FAProp(10.,5.5)(10.,10.)(0.,){/Sine}{1}
\FALabel(8.93,7.75)[r]{$W$}
\FAVert(10.,14.5){0}
\FAVert(10.,5.5){0}
\FAVert(10.,10.){1}
\end{feynartspicture}
\end{small}
\vspace{-4em}
\caption{
Corrections to the $WWH$ vertex and the corresponding counter-term
diagram.  The label $\xtilde f$ denotes all (s)fermions, except in the
presence of a $\xtilde f\,\,'$, in which case the former denotes only the
isospin-up and the latter the isospin-down members of the (s)fermion
doublets.
}
\label{fig:WWhHvert}
\end{center}
 \vspace{-3em}
\end{figure}

\begin{figure}[ht!]
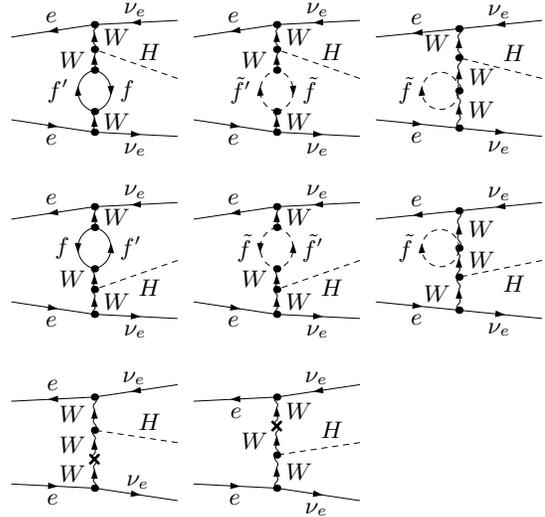

\vspace{-4em}
\begin{center}
\begin{small}
\unitlength=.68bp%
\begin{feynartspicture}(324,303)(3,3)
\FADiagram{}
\FAProp(0.,15.)(10.,16.5)(0.,){/Straight}{-1}
\FALabel(4.77007,16.8029)[b]{$e$}
\FAProp(0.,5.)(10.,3.5)(0.,){/Straight}{1}
\FALabel(4.77007,3.19715)[t]{$e$}
\FAProp(20.,17.)(10.,16.5)(0.,){/Straight}{1}
\FALabel(14.9226,17.8181)[b]{$\nu_e$}
\FAProp(20.,10.)(10.,13.5)(0.,){/ScalarDash}{0}
\FALabel(15.3153,12.17)[lb]{$H$}
\FAProp(20.,3.)(10.,3.5)(0.,){/Straight}{-1}
\FALabel(14.9226,2.18193)[t]{$\nu_e$}
\FAProp(10.,16.5)(10.,13.5)(0.,){/Sine}{-1}
\FALabel(11.07,15.)[l]{$W$}
\FAProp(10.,3.5)(10.,6.)(0.,){/Sine}{1}
\FALabel(11.07,4.75)[l]{$W$}
\FAProp(10.,13.5)(10.,11.)(0.,){/Sine}{-1}
\FALabel(8.93,12.25)[r]{$W$}
\FAProp(10.,6.)(10.,11.)(0.8,){/Straight}{-1}
\FALabel(13.07,8.5)[l]{$f$}
\FAProp(10.,6.)(10.,11.)(-0.8,){/Straight}{1}
\FALabel(6.93,8.5)[r]{$f'$}
\FAVert(10.,16.5){0}
\FAVert(10.,3.5){0}
\FAVert(10.,13.5){0}
\FAVert(10.,6.){0}
\FAVert(10.,11.){0}

\FADiagram{}
\FAProp(0.,15.)(10.,16.5)(0.,){/Straight}{-1}
\FALabel(4.77007,16.8029)[b]{$e$}
\FAProp(0.,5.)(10.,3.5)(0.,){/Straight}{1}
\FALabel(4.77007,3.19715)[t]{$e$}
\FAProp(20.,17.)(10.,16.5)(0.,){/Straight}{1}
\FALabel(14.9226,17.8181)[b]{$\nu_e$}
\FAProp(20.,10.)(10.,13.5)(0.,){/ScalarDash}{0}
\FALabel(15.3153,12.17)[lb]{$H$}
\FAProp(20.,3.)(10.,3.5)(0.,){/Straight}{-1}
\FALabel(14.9226,2.18193)[t]{$\nu_e$}
\FAProp(10.,16.5)(10.,13.5)(0.,){/Sine}{-1}
\FALabel(11.07,15.)[l]{$W$}
\FAProp(10.,3.5)(10.,6.)(0.,){/Sine}{1}
\FALabel(11.07,4.75)[l]{$W$}
\FAProp(10.,13.5)(10.,11.)(0.,){/Sine}{-1}
\FALabel(8.93,12.25)[r]{$W$}
\FAProp(10.,6.)(10.,11.)(0.8,){/ScalarDash}{-1}
\FALabel(13.07,8.5)[l]{$\tilde f$}
\FAProp(10.,6.)(10.,11.)(-0.8,){/ScalarDash}{1}
\FALabel(6.93,8.5)[r]{$\tilde f'$}
\FAVert(10.,16.5){0}
\FAVert(10.,3.5){0}
\FAVert(10.,13.5){0}
\FAVert(10.,6.){0}
\FAVert(10.,11.){0}

\FADiagram{}
\FAProp(0.,15.)(10.,16.)(0.,){/Straight}{-1}
\FALabel(4.84577,16.5623)[b]{$e$}
\FAProp(0.,5.)(10.,4.)(0.,){/Straight}{1}
\FALabel(4.84577,3.43769)[t]{$e$}
\FAProp(20.,17.)(10.,16.)(0.,){/Straight}{1}
\FALabel(14.8458,17.5623)[b]{$\nu_e$}
\FAProp(20.,10.)(10.,12.5)(0.,){/ScalarDash}{0}
\FALabel(15.3153,12.0312)[lb]{$H$}
\FAProp(20.,3.)(10.,4.)(0.,){/Straight}{-1}
\FALabel(14.8458,2.43769)[t]{$\nu_e$}
\FAProp(10.,16.)(10.,12.5)(0.,){/Sine}{-1}
\FALabel(8.93,14.25)[r]{$W$}
\FAProp(10.,4.)(10.,8.5)(0.,){/Sine}{1}
\FALabel(11.07,6.25)[l]{$W$}
\FAProp(10.,12.5)(10.,8.5)(0.,){/Sine}{-1}
\FALabel(11.12,10.15)[l]{$W$}
\FAProp(10.,8.5)(10.,8.5)(5.5,8.5){/ScalarDash}{-1}
\FALabel(4.43,8.5)[r]{$\tilde f$}
\FAVert(10.,16.){0}
\FAVert(10.,4.){0}
\FAVert(10.,12.5){0}
\FAVert(10.,8.5){0}

\FADiagram{}
\FAProp(0.,15.)(10.,16.5)(0.,){/Straight}{-1}
\FALabel(4.77007,16.8029)[b]{$e$}
\FAProp(0.,5.)(10.,3.5)(0.,){/Straight}{1}
\FALabel(4.77007,3.19715)[t]{$e$}
\FAProp(20.,17.)(10.,16.5)(0.,){/Straight}{1}
\FALabel(14.9226,17.8181)[b]{$\nu_e$}
\FAProp(20.,10.)(10.,6.5)(0.,){/ScalarDash}{0}
\FALabel(15.3153,7.68)[lt]{$H$}
\FAProp(20.,3.)(10.,3.5)(0.,){/Straight}{-1}
\FALabel(14.9226,2.18193)[t]{$\nu_e$}
\FAProp(10.,16.5)(10.,14.)(0.,){/Sine}{-1}
\FALabel(11.07,15.25)[l]{$W$}
\FAProp(10.,3.5)(10.,6.5)(0.,){/Sine}{1}
\FALabel(11.07,5.)[l]{$W$}
\FAProp(10.,6.5)(10.,9.)(0.,){/Sine}{1}
\FALabel(8.93,7.75)[r]{$W$}
\FAProp(10.,14.)(10.,9.)(0.8,){/Straight}{1}
\FALabel(6.93,11.5)[r]{$f$}
\FAProp(10.,14.)(10.,9.)(-0.8,){/Straight}{-1}
\FALabel(13.07,11.5)[l]{$f'$}
\FAVert(10.,16.5){0}
\FAVert(10.,3.5){0}
\FAVert(10.,6.5){0}
\FAVert(10.,14.){0}
\FAVert(10.,9.){0}

\FADiagram{}
\FAProp(0.,15.)(10.,16.5)(0.,){/Straight}{-1}
\FALabel(4.77007,16.8029)[b]{$e$}
\FAProp(0.,5.)(10.,3.5)(0.,){/Straight}{1}
\FALabel(4.77007,3.19715)[t]{$e$}
\FAProp(20.,17.)(10.,16.5)(0.,){/Straight}{1}
\FALabel(14.9226,17.8181)[b]{$\nu_e$}
\FAProp(20.,10.)(10.,6.5)(0.,){/ScalarDash}{0}
\FALabel(15.3153,7.68)[lt]{$H$}
\FAProp(20.,3.)(10.,3.5)(0.,){/Straight}{-1}
\FALabel(14.9226,2.18193)[t]{$\nu_e$}
\FAProp(10.,16.5)(10.,14.)(0.,){/Sine}{-1}
\FALabel(11.07,15.25)[l]{$W$}
\FAProp(10.,3.5)(10.,6.5)(0.,){/Sine}{1}
\FALabel(11.07,5.)[l]{$W$}
\FAProp(10.,6.5)(10.,9.)(0.,){/Sine}{1}
\FALabel(8.93,7.75)[r]{$W$}
\FAProp(10.,14.)(10.,9.)(0.8,){/ScalarDash}{1}
\FALabel(6.93,11.5)[r]{$\tilde f$}
\FAProp(10.,14.)(10.,9.)(-0.8,){/ScalarDash}{-1}
\FALabel(13.07,11.5)[l]{$\tilde f'$}
\FAVert(10.,16.5){0}
\FAVert(10.,3.5){0}
\FAVert(10.,6.5){0}
\FAVert(10.,14.){0}
\FAVert(10.,9.){0}

\FADiagram{}
\FAProp(0.,15.)(10.,16.)(0.,){/Straight}{-1}
\FALabel(4.84577,16.5623)[b]{$e$}
\FAProp(0.,5.)(10.,4.)(0.,){/Straight}{1}
\FALabel(4.84577,3.43769)[t]{$e$}
\FAProp(20.,17.)(10.,16.)(0.,){/Straight}{1}
\FALabel(14.8458,17.5623)[b]{$\nu_e$}
\FAProp(20.,10.)(10.,8.)(0.,){/ScalarDash}{0}
\FALabel(15.3153,8.20525)[lt]{$H$}
\FAProp(20.,3.)(10.,4.)(0.,){/Straight}{-1}
\FALabel(14.8458,2.43769)[t]{$\nu_e$}
\FAProp(10.,16.)(10.,11.5)(0.,){/Sine}{-1}
\FALabel(11.07,13.75)[l]{$W$}
\FAProp(10.,4.)(10.,8.)(0.,){/Sine}{1}
\FALabel(8.93,6.)[r]{$W$}
\FAProp(10.,8.)(10.,11.5)(0.,){/Sine}{1}
\FALabel(11.12,10.1)[l]{$W$}
\FAProp(10.,11.5)(10.,11.5)(5.5,11.5){/ScalarDash}{-1}
\FALabel(4.43,11.5)[r]{$\tilde f$}
\FAVert(10.,16.){0}
\FAVert(10.,4.){0}
\FAVert(10.,8.){0}
\FAVert(10.,11.5){0}

\FADiagram{}
\FAProp(0.,15.)(10.,15.5)(0.,){/Straight}{-1}
\FALabel(4.9226,16.3181)[b]{$e$}
\FAProp(0.,5.)(10.,4.5)(0.,){/Straight}{1}
\FALabel(4.9226,3.68193)[t]{$e$}
\FAProp(20.,17.)(10.,15.5)(0.,){/Straight}{1}
\FALabel(14.7701,17.3029)[b]{$\nu_e$}
\FAProp(20.,10.)(10.,11.5)(0.,){/ScalarDash}{0}
\FALabel(15.3153,11.5556)[lb]{$H$}
\FAProp(20.,3.)(10.,4.5)(0.,){/Straight}{-1}
\FALabel(15.2299,4.80285)[b]{$\nu_e$}
\FAProp(10.,8.)(10.,4.5)(0.,){/Sine}{-1}
\FALabel(8.93,6.25)[r]{$W$}
\FAProp(10.,8.)(10.,11.5)(0.,){/Sine}{1}
\FALabel(8.93,9.75)[r]{$W$}
\FAProp(10.,15.5)(10.,11.5)(0.,){/Sine}{-1}
\FALabel(8.93,13.5)[r]{$W$}
\FAVert(10.,15.5){0}
\FAVert(10.,4.5){0}
\FAVert(10.,11.5){0}
\FAVert(10.,8.){1}

\FADiagram{}
\FAProp(0.,15.)(10.,15.5)(0.,){/Straight}{-1}
\FALabel(5.0774,14.1819)[t]{$e$}
\FAProp(0.,5.)(10.,4.5)(0.,){/Straight}{1}
\FALabel(4.9226,3.68193)[t]{$e$}
\FAProp(20.,17.)(10.,15.5)(0.,){/Straight}{1}
\FALabel(14.7701,17.3029)[b]{$\nu_e$}
\FAProp(20.,10.)(10.,8.5)(0.,){/ScalarDash}{0}
\FALabel(15.3153,10.556)[lb]{$H$}
\FAProp(20.,3.)(10.,4.5)(0.,){/Straight}{-1}
\FALabel(14.7701,2.69715)[t]{$\nu_e$}
\FAProp(10.,12.)(10.,15.5)(0.,){/Sine}{1}
\FALabel(11.07,13.75)[l]{$W$}
\FAProp(10.,12.)(10.,8.5)(0.,){/Sine}{-1}
\FALabel(8.93,10.25)[r]{$W$}
\FAProp(10.,4.5)(10.,8.5)(0.,){/Sine}{1}
\FALabel(11.07,6.5)[l]{$W$}
\FAVert(10.,15.5){0}
\FAVert(10.,4.5){0}
\FAVert(10.,8.5){0}
\FAVert(10.,12.){1}
\end{feynartspicture}
\end{small}
\vspace{-4em}
\caption{
Corrections to the $W$-boson propagator and the corresponding counter-term
diagrams.  The labeling is as in \reffi{fig:WWhHvert}.
}
\label{fig:WWhHself}
\end{center}
\vspace{-2em}
\end{figure}

\begin{figure}[ht!]
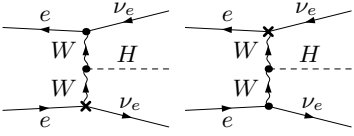

\vspace{-3em}
\begin{center}
\begin{small}
\unitlength=.68bp%
\begin{feynartspicture}(324,101)(3,1)
\FADiagram{}
\FAProp(0.,15.)(10.,14.5)(0.,){/Straight}{-1}
\FALabel(5.0774,15.8181)[b]{$e$}
\FAProp(0.,5.)(10.,5.5)(0.,){/Straight}{1}
\FALabel(5.0774,4.18193)[t]{$e$}
\FAProp(20.,17.)(10.,14.5)(0.,){/Straight}{1}
\FALabel(14.6241,16.7737)[b]{$\nu_e$}
\FAProp(20.,10.)(10.,10.)(0.,){/ScalarDash}{0}
\FALabel(15.,10.82)[b]{$H$}
\FAProp(20.,3.)(10.,5.5)(0.,){/Straight}{-1}
\FALabel(15.3759,5.27372)[b]{$\nu_e$}
\FAProp(10.,14.5)(10.,10.)(0.,){/Sine}{-1}
\FALabel(8.93,12.25)[r]{$W$}
\FAProp(10.,5.5)(10.,10.)(0.,){/Sine}{1}
\FALabel(8.93,7.75)[r]{$W$}
\FAVert(10.,14.5){0}
\FAVert(10.,10.){0}
\FAVert(10.,5.5){1}

\FADiagram{}
\FAProp(0.,15.)(10.,14.5)(0.,){/Straight}{-1}
\FALabel(5.0774,15.8181)[b]{$e$}
\FAProp(0.,5.)(10.,5.5)(0.,){/Straight}{1}
\FALabel(5.0774,4.18193)[t]{$e$}
\FAProp(20.,17.)(10.,14.5)(0.,){/Straight}{1}
\FALabel(14.6241,16.7737)[b]{$\nu_e$}
\FAProp(20.,10.)(10.,10.)(0.,){/ScalarDash}{0}
\FALabel(15.,10.82)[b]{$H$}
\FAProp(20.,3.)(10.,5.5)(0.,){/Straight}{-1}
\FALabel(15.3759,5.27372)[b]{$\nu_e$}
\FAProp(10.,14.5)(10.,10.)(0.,){/Sine}{-1}
\FALabel(8.93,12.25)[r]{$W$}
\FAProp(10.,5.5)(10.,10.)(0.,){/Sine}{1}
\FALabel(8.93,7.75)[r]{$W$}
\FAVert(10.,5.5){0}
\FAVert(10.,10.){0}
\FAVert(10.,14.5){1}
\end{feynartspicture}
\end{small}
\vspace{-4em}
\caption{
Counter-term contributions entering via the $e\,\nu_e\,W$ vertex.
}
\label{fig:enuWCT}
\end{center}
\vspace{-3em}
\end{figure}

Below we describe the loop corrections that enter the process \eennH\
at the \onel\ level%
\footnote{%
We focus on the description of the loop corrections to the $WW$-fusion
channel. However, the same corrections are taken into account for the
Higgs-strahlung process.}%
.
The contributions involve corrections to the $WWH$ vertex and the
corresponding counter-term diagram, see \reffi{fig:WWhHvert},
corrections to the $W$-boson propagators and the corresponding counter
terms, see \reffi{fig:WWhHself}, and the counter-term contributions
to the $e\nu_e W$ vertex, see \reffi{fig:enuWCT}. Furthermore,
Higgs propagator corrections enter via the wave-function normalization of
the external Higgs boson, see below. There are
also $W$-boson propagator corrections inducing a transition from the
$W^\pm$ to either $G^\pm$ or $H^\pm$.  These corrections affect only the
longitudinal part of the $W$~boson and are thus $\propto
m_e/\MW$ and have been neglected.

While the renormalization in the counter terms depicted in
\reffis{fig:WWhHself} and \ref{fig:enuWCT} is as in the SM,
the $WWH$ vertex is renormalized as follows,
\BE
\begin{aligned}
\Ga&\mbox{}_{WWH}^{(0), {\rm CT}} = \GHn 
  \Biggl[ 1 + \de \tilde Z_e 
            + \edz \frac{\de \MW^2}{\MW^2}
            + \de Z_W \\
&{}         + \frac{\de \sw}{\sw} \notag 
            - \sinb\Cb \frac{\Sba}{\Cba} \de \tb \\
&           + \edz \de Z_H
            + \edz \frac{\Ghn}{\GHn} \de Z_{hH}
  \Biggr]\,,
\label{eq:WWHct}
\end{aligned}
\EE
with $\Ga_{\{h,H\}}^{(0)} = i e \MW/\sw \{\sin,\cos\}(\be - \al)$.
The counter-terms are given by
\BEA
\de Z_{\{H,h\}} &=& 
 -\KKL \re\Si_{\{HH,hh\}}'(m_{\{H,h\}}^2) \KKR^{\rm div}, \non \\
\de Z_{hH} &=& \de Z_{Hh} = 
               \frac{\Sa\Ca}{\CZa} (\de Z_{h} - \de Z_{H})\,, \non \\
\de\tb &=& \de\tb^{\msbarm}  \; = \; - \ed{2\CZa} \times \non \\
&&  \KKL \re\Si_{hh}'(\mh^2) - \re\Si_{HH}'(\mH^2) \KKR^{\rm div}\,, \non \\
\de M_V^2 &=& \re\Si_V^T(M_V^2)\,, (V = Z, W^\pm)\, , \non \\
\frac{\de\sw}{\sw} &=& \edz \frac{\cw^2}{\sw^2}
  \KL \frac{\de\MZ^2}{\MZ^2} - \frac{\de\MW^2}{\MW^2} \KR, \non \\
\de \tilde Z_e &=& \de Z_e - \edz \De r, \non \\
       &=& \edz \Big\{ \frac{\cw^2}{\sw^2} 
       \KL \frac{\de\MZ^2}{\MZ^2} - \frac{\de\MW^2}{\MW^2} \KR \non \\
       && - \KL \Si_W^T(0) - \de\MW^2 \KR/\MW^2
                \Big\} \,.
\label{eq:rcs}
\EEA
$\de \tilde Z_e$ incorporates the charge renormalization and
$\De r$ (which is due to the parametrization of the Born cross section
with $G_\mu$), see~\cite{eennH}.
The gauge-boson field-renormalization constants,
$\de Z_{\{W,Z,\ga Z\}}$, drop out in the result for the
complete S-matrix element.

In order to ensure the correct on-shell properties of the outgoing
Higgs boson, which are necessary for the correct normalization of 
the S-matrix element, furthermore a finite wave-function normalization 
has to be incorporated~\cite{hff}:
\BE
\Ga^{{\rm WF}} = \GHn 
  \Biggl[ \sqrt{\rZH} - 1
                 + \edz \frac{\Ghn}{\GHn} \sqrt{\rZH} \; \rZhH 
  \Biggr]\,.
\label{eq:WWHct2}
\EE
The finite wave-function renormalizations are given in \citere{eennH}.

The evaluation of the Feynman diagrams and the further calculations
have been done using the packages {\em FeynArts}, {\em FormCalc}, and 
{\em LoopTools}~\cite{feynarts,formcalc,fa-fc-lt}.
The Higgs-boson sector evaluations have been performed with
\fh2.0~\cite{feynhiggs,fhlatest}. The resulting Fortran code will be
made available at {\tt www.hep-processes.de}.


\section{NUMERICAL ANALYSIS}

The numerical analysis is performed in the ``$\siHenh$'' (``enhanced
cross section'') scenario, which is defined by
\BEA
\label{xsmax}
&& \mt = 174.3 \gev,\; \msusy = 350 \gev, \non \\
&& \Xt = 2\msusy,\; \Ab = \At, \mgl = 800 \gev \non \\
&& \mu = 1000 \gev, M_2 = 200 \gev~, 
\EEA
which, up to changes in the values for $\msusy$ and $\mu$, is the
well-known $\mhmax$~scenario~\cite{LHbenchmark}. 

\begin{figure}[htb]
\vspace{-2em}
\includegraphics[width=7.5cm]{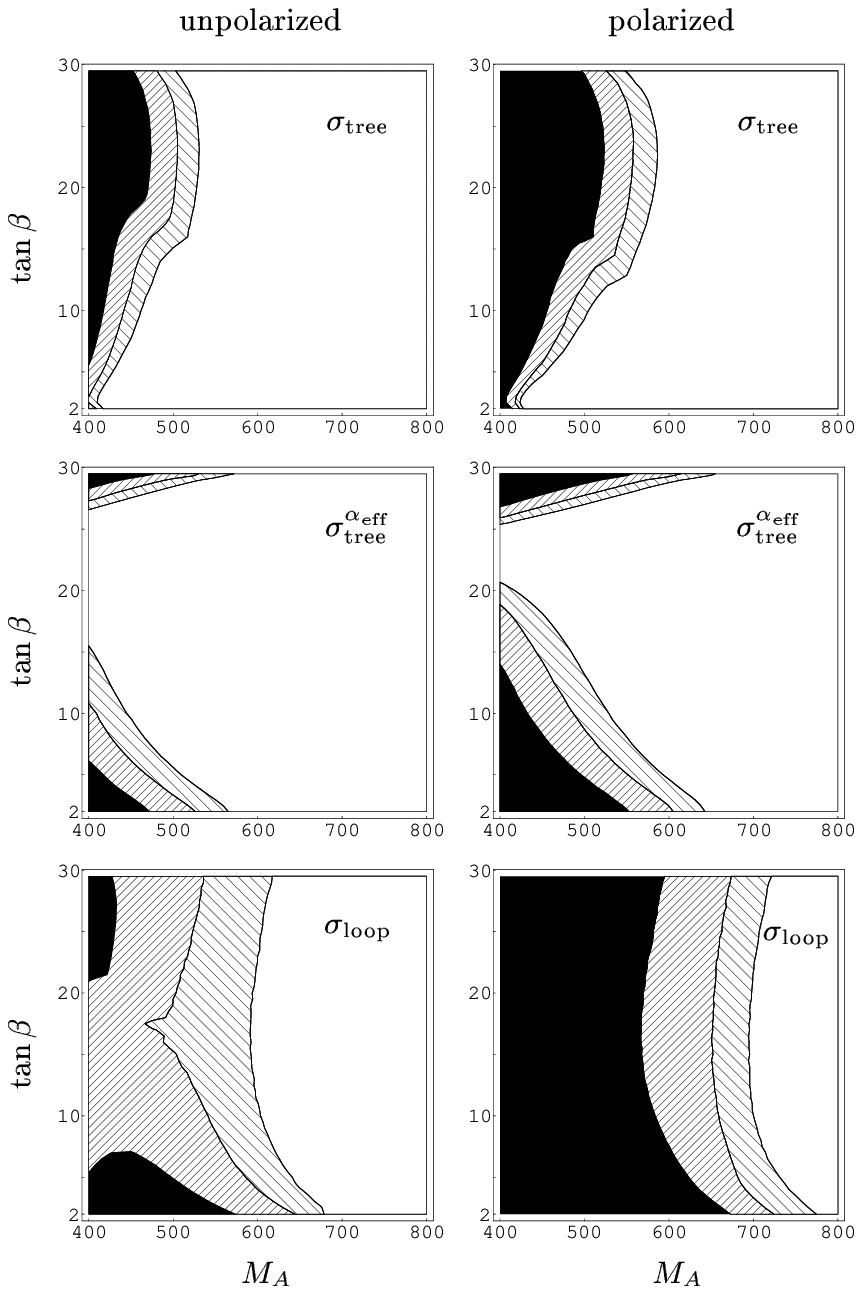}
\vspace{-4em}
\caption{The cross section for \eennH\ is shown in the
  $\siHenh$~scenario in the $\MA-\tb$-plane for $\sqrt{s} = 1 \tev$.
  The different shadings correspond to:
white: $\si \le 0.01 \fb$, light shaded: $0.01 \fb \le \si \le 0.02 \fb$, 
dark shaded: $0.02 \fb \le \si \le 0.05 \fb$, black: $\si \ge 0.05 \fb$.}
\label{fig:siHenh}
\vspace{-2.5em}
\end{figure}

In \reffi{fig:siHenh} we show the result for the $\si_H$ evaluation in
the $\siHenh$~scenario for $\sqrt{s} = 1 \tev$. 
Assuming an integrated luminosity of \order{1-2\,\iabm}, a cross section
of $\si_H = 0.01 \fb$ constitutes a lower limit for which the
observation of the heavy Higgs boson could be possible.
We first focus on the
left column, where the 
results are given without polarization of the $e^+$ and $e^-$ beams. 
The upper row shows the tree-level cross
section (including the finite wave-function renormalization,
\refeq{eq:WWHct2}). The middle row shows the 
$\aeff$~approximation for $\si_H$ (which has mostly been used for
phenomenological analyses so far), where the outgoing Higgs boson is
not on-shell. The lower row presents the
result including the \onel\ corrections. Two conclusions can be drawn:
neglecting the wave function renormalization leads to considerable
differences to the case in which the outgoing Higgs boson is
on-shell. Secondly, the loop corrections, 
\reffis{fig:WWhHvert} - \ref{fig:enuWCT}, lead to a strong enhancement
of the $H$ production cross section. Instead of 
$\MH \lsim \sqrt{s}/2 = 500 \gev$, now the observation up to 
$\MH \approx 600 \gev$ seems to be possible.

The prospects for observing a heavy Higgs boson beyond the kinematical
limit of $\MH \lsim \sqrt{s}/2$ become even more favorable 
if polarized beams are used. 
The cross section becomes enhanced for left-handedly polarized~$e^-$
and right-handedly polarized~$e^+$. While a 100\%
polarization results in a cross section enhancement of roughly a
factor of 4, more realistic values of 80\% polarization for~$e^-$
and 60\% polarization for~$e^+$~\cite{polarization}
would yield roughly an enhancement by a factor of 3. 
The right column of \reffi{fig:siHenh} shows the $\siHenh$~scenario
with 100\% polarization of both beams. The area in the  
$\MA-\tb$ plane in which observation of the $H$~boson might become
possible is strongly increased in this case. 
Thus, in the case with beam polarization, taking into account the
\onel\ corrections, $\MH \lsim 750 \gev$ could be observable at 
$\sqrt{s} = 1 \tev$.

\begin{figure}[!htb]
\vspace{-2em}
\includegraphics[width=7.5cm]{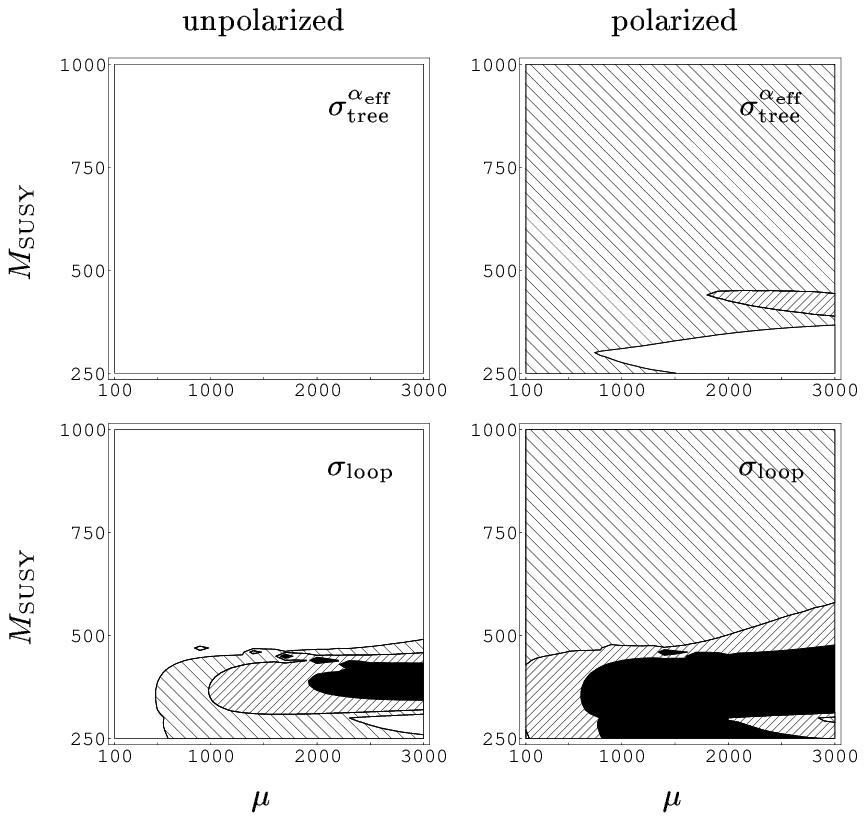}
\vspace{-4em}
\caption{The cross section for \eennH\ is shown in the
  $\mhmax$~scenario for $\MA = 600 \gev$
  and $\tb = 4$ in the $\mu-\msusy$ plane for $\sqrt{s} = 1 \tev$. The
  color coding is as in \reffi{fig:siHenh}. }
\label{fig:MA600tb4}
\vspace{-2em}
\end{figure}

\begin{figure}[htb]
\vspace{-2em}
\includegraphics[width=7.5cm]{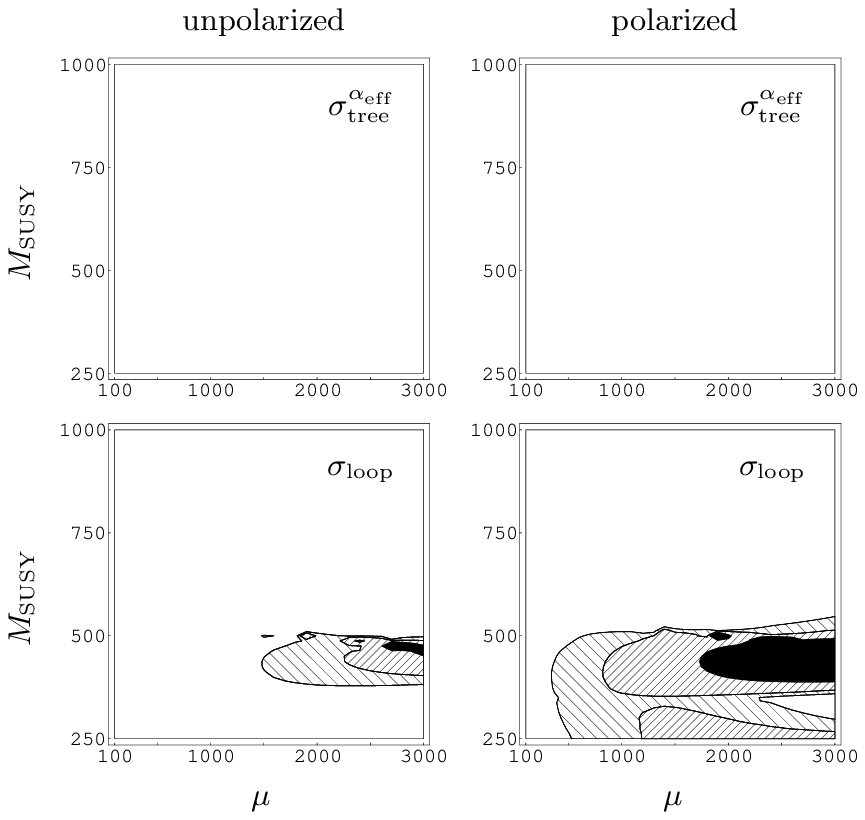}
\vspace{-4em}
\caption{The cross section for \eennH\ is shown in the
  $\mhmax$~scenario for $\MA = 700 \gev$
  and $\tb = 4$ in the $\mu-\msusy$ plane for $\sqrt{s} = 1 \tev$. The
  color coding is as in \reffi{fig:siHenh}. }
\label{fig:MA700tb4}
\vspace{-2em}
\end{figure}

The question whether this enlarged reach in $\MH$ is due to a special
choice of MSSM parameters is analyzed in \reffis{fig:MA600tb4},
\ref{fig:MA700tb4}. We show the results for $\si_H$ in the
$\mu-\msusy$-plane for $\MA$ fixed to $\MA = 600, 700 \gev$
(\reffi{fig:MA600tb4}, \ref{fig:MA700tb4}) and $\tb = 4$ with the
other parameters chosen as in the 
$\mhmax$~scenario. The $\aeff$~approximation
is compared with the \onel\ result (including also the finite
wave-function renormalization). In the case of unpolarized beams (left
columns) the $\aeff$~result shows no region of
observability in the $\mu-\msusy$-plane for $\MA = 600, 700 \gev$. 
However, taking the loop corrections into account, $\si_H$ becomes
large enough to observe \eennH\ in a sizable fraction of the
$\mu-\msusy$-plane (with $\msusy \lsim 500 \gev$). The situation becomes
even more favorable if besides the loop correction also polarization
is taken into account. For $\MA = 600 \gev$ (\reffi{fig:MA600tb4}) the
whole $\mu-\msusy$-plane possesses an observable $\si_H$, even with
$\si_H > 0.05 \fb$ for $\msusy \lsim 500 \gev$. For 
$\MA = 700 \gev$, in the case of polarized beams, loop corrections
enhance $\si_H$ to an observable level for $\msusy \lsim 500 \gev$ for
nearly all $\mu$ values. 
Thus, an enhanced cross section, although clearly dependent on the
chosen scenario, can be found in large parts of the MSSM parameter space. 
%
%
%


\section{CONCLUSIONS}

We have investigated the production of the heavy $\cp$-even MSSM
Higgs-boson at a future LC in the process \eennH, which
is dominated by the $WW$-fusion mechanism at higher energies.
We have
evaluated all one-loop contributions from fermions and sfermions, and we
have implemented the numerically large process-independent
Higgs-boson propagator corrections so that the correct
on-shell properties of the outgoing Higgs boson are ensured.
We find that in favorable regions of the MSSM parameter space the genuine 
loop corrections can drastically enlarge the parameter space for which
detection of $H$ becomes possible. In such
a scenario, assuming polarized beams, at $\sqrt{s} = 1 \tev$ the
detection of $H$~could be possible up to $\MH \lsim 750 \gev$.  

\smallskip
\noindent
{\em Acknowledgements:} S.H. thanks the organizers of ``RADCOR 2002 -- 
Loops \& Legs 2002'' for the invitation and the inspiring atmosphere.



\end{document}